\documentclass[reprint,preprintnumbers,
amsmath,amssymb,aps,nofootinbib,superscriptaddress]{revtex4-1}
\usepackage[pdftex]{graphicx}
\usepackage{dcolumn}
\usepackage{bm}
\usepackage{natbib}
\usepackage{cases, comment, subcaption}
\usepackage[pdftex]{hyperref}
\usepackage{color}
\definecolor{indigo(dye)}{rgb}{0.0, 0.25, 0.42}
\definecolor{phthaloblue}{rgb}{0.0, 0.06, 0.54}
\hypersetup{
    colorlinks=true,
    linkcolor=phthaloblue,
    citecolor=phthaloblue,
    filecolor=phthaloblue,
    urlcolor=phthaloblue,
    }
\usepackage[utf8]{inputenc}

\begin{document}
\preprint{\begin{minipage}[b]{1\linewidth}
\begin{flushright}TU-1087 \\
IPMU19-0060\\
KEK-TH-2127\\
KEK-Cosmo-235
\end{flushright}
\end{minipage}}
\title{Bouncing Universe from Nothing}
\author{Hiroki Matsui}
\email{hiroki.matsui.c6@tohoku.ac.jp}
\affiliation{Department of Physics, Tohoku University, Sendai, 980-8578 Japan}
\author{Fuminobu Takahashi}
\email{fumi@tohoku.ac.jp}
\affiliation{Department of Physics, Tohoku University, Sendai, 980-8578 Japan}
\affiliation{Kavli IPMU (WPI), UTIAS, The University of Tokyo, Kashiwa, Chiba 277-8583, Japan}
\author{Takahiro Terada}
\email{teradat@post.kek.jp}
\affiliation{Theory Center, IPNS, KEK, 1-1 Oho, Tsukuba, Ibaraki 305-0801, Japan}

\begin{abstract}
We find a class of solutions for a homogeneous and isotropic universe in which the initially expanding universe stops expanding,  experiences contraction, and then expands again (the ``bounce"),  in the framework of Einstein gravity with a real scalar field without violating the null energy condition nor encountering any singularities.
Two essential ingredients for the bouncing universe are the positive spatial curvature and the scalar potential which becomes flatter at large field values. Depending on the initial condition, either the positive curvature or the negative potential stops the cosmic expansion and begins the contraction phase. The flat potential plays a crucial role in triggering the bounce.  After the bounce, the flat potential naturally allows the universe to enter the slow-roll inflation regime, thereby making the bouncing universe compatible with observations. If the $e$-folding of the subsequent inflation is just enough, a positive spatial curvature may be found in the future observations. Our scenario nicely fits with the creation of the universe from nothing, which leads to the homogeneous and isotropic universe with positive curvature.  As a variant of the mechanism, we also find solutions representing a cyclic universe. 
\end{abstract}
\date{\today}
\maketitle
\flushbottom
\allowdisplaybreaks[1]

\section{Introduction}
The  $\Lambda$CDM model based on the slow-roll inflation is extremely successful, and observations of the cosmic microwave background (CMB) have determined various cosmological parameters with a high precision~\cite{Aghanim:2018eyx}.  Despite its great success,  there remain some unanswered questions. One of them is the initial condition of the universe: we still do not know how the universe began before inflation. This is due to the attractor nature of inflation. Once the inflation starts,  information on the initial condition is soon forgotten. In particular, the initial spatial curvature is stretched away by the exponential expansion of inflation. 

The expanding universe is consistent with all the observations, and it is taken for granted that the universe keeps expanding from the very beginning until the present. Indeed, the Friedmann equation tells us that a flat expanding universe keeps expanding as long as the total energy of the universe is positive.  However, it is possible, at least theoretically, that the universe experienced a contraction phase in the very early epoch before inflation or it may enter the contraction phase in the far future. 

It is either a positive spatial curvature or a negative potential that can cause the contraction of the universe. As is well known, a closed (positively curved) universe stops expanding and starts to contract at a certain point.  Also, scalar potentials which become negative for a range of field values  (for brevity we call them negative potentials) can lead to a contracting phase when the Hubble parameter $H(t)$ passes through zero and changes its sign.  Cosmology with a negative potential has been extensively studied in Refs.~\cite{Linde:2001ae, Felder:2002jk}.  

A negative potential appears in various situations. For example, the potential of the Standard Model Higgs may become negative around the true minimum depending on the precise value of the top quark mass~\cite{Degrassi:2012ry, Buttazzo:2013uya}.
In supergravity, the scalar potential receives a negative contribution from spontaneous breaking of  $R$-symmetry, 
and the potential is always non-positive for supersymmetric solutions. Therefore, depending on the uplift by supersymmetry breaking effect,  some part of the potential may remain negative.
 In the string theory landscape, we expect there are many local vacua with a negative potential~\cite{Douglas:2006es}.

Once the universe starts to contract,  the kinetic energy of scalar fields increases for generic potential.  Once the kinetic energy dominates the universe, it is hard to stop the universe from collapsing to a singularity\footnote{Of course, detailed knowledge of quantum gravity is required to describe the evolution of the universe when the energy density reaches the Planck scale.}, as the kinetic energy rapidly increases as $a^{-6}$ where $a$ is the scale factor.

Such a catastrophic collapse can be avoided if the universe stops contracting and starts to expand again, the so-called bouncing universe.  If the bounce takes place before the energy reaches the Planck scale,  one can reliably use Einstein gravity as an effective field theory.  As we shall see shortly, the existence of the bouncing solution depends on the spatial curvature. For a flat or open (negatively curved) universe, the possibility of a bounce is precluded by the null energy condition (NEC).  For a closed universe, on the other hand, the bounce takes place when the curvature term balances with the total energy of the universe. However, it is challenging to satisfy this condition since the curvature term grows as $a^{-2}$ while the kinetic energy of the scalar field tends to grow much more rapidly.

In this paper, we find a class of solutions for a Friedmann-Lema\^{i}tre-Robertson-Walker (FLRW) universe in which the initially expanding universe begins to contract and finally expands again,  in the framework of Einstein gravity with a real scalar field.  Two essential ingredients we introduce are the positive spatial curvature and the scalar potential that becomes flatter as the scalar field goes away from the minimum.
The transition from expansion to contraction is induced either by the spatial curvature or the negative potential. After the scalar field passes the potential minimum, it continues to climb up the potential, converting most of the kinetic energy to the potential energy. In particular, if the potential becomes sufficiently flat when the kinetic energy
gets suppressed,  the scalar field slow-rolls on a flat plateau of the potential, and the universe experiences an accelerated ($\ddot{a}>0$) contraction. If the positive curvature term balances with the scalar field energy during this period,  the bounce occurs.  We emphasize that the accelerated contraction phase induced by the flat potential plays a crucial role for the bouncing universe. 

After the bounce,  the scalar field still remains on a flat plateau of the potential, and so, the universe naturally experiences inflation. It is interesting that the flat potential required for the bounce results in the subsequent slow-roll inflation. If the scalar field continues to slow-roll for a sufficiently long time and finally finds another minimum with a sufficiently small and positive cosmological constant,  our bouncing solution can be consistent with observations. In particular, we may be able to see the effects of the contraction phase and/or the positive curvature if the $e$-folding number of inflation is just enough.

Interestingly, positive curvature is weakly preferred by the latest CMB observation~\cite{Aghanim:2018eyx}, although it is fair to say that it is currently consistent with the flat universe when the lensing and baryon acoustic oscillation (BAO) constraints are combined.  If future observations confirm positive curvature, it might be the remnant of the fact that our universe experienced a contraction phase in the very early universe.

Lastly, let us mention the related works in the past.
A negative cosmological constant and positive spatial curvature were employed in Refs.~\cite{1990MNRAS.243..252K, Dabrowski:1995ae, Graham:2011nb, Graham:2014pca} to realize a cyclic universe.\footnote{ 
 See also Refs.~\cite{Biswas:2008ti, Biswas:2008kj, Barrow:2017yqt, Ganguly:2017qff} for other models of the cyclic universe.
}  Instead of a scalar field, they utilized domain walls to satisfy the necessary condition for the bounce to occur. Note that  one can construct a non-singular bouncing or cyclic universe with a negative scalar potential instead of the domain walls~\cite{Dabrowski:1995ae}, 
because, as a mathematical procedure, it is possible to reconstruct a scalar potential that reproduces any given solution of the scale factor as a function of time,  as long as $\mathcal{K}/a^2 - \dot{H} = \dot{\phi}^2/2$ is not negative~\cite{Martin:2003sf}.
In contrast, we begin with a scalar field model with some reasonable choices of the scalar potential which is well-motivated in the context of inflation model building, and study how the bounce occurs.

We emphasize here that it is far from trivial to build a model in which the universe is initially expanding, then contracts, and finally expands again, although each step was relatively well studied 
 (see Refs.~\cite{Linde:2001ae, Felder:2002jk} for the turnaround to contraction by a negative potential and Refs.~\cite{ Gordon:2002jw, Martin:2003sf, Falciano:2008gt, deHaro:2015yyh, Parker:1973qd, 1978SvAL....4...82S, Barrow:1980en, HawkingLecture, Page:1984qt, Schmidt:1990yg, Cornish:1997ah, Kamenshchik:1998ue} for the bounce with positive curvature)
 There are many attempts to realize bouncing or cyclic cosmology by violating the NEC~\cite{delaMacorra:2004et,Buchbinder:2007ad, Graham:2017hfr}, but we do not consider such a possibility throughout the paper.  Also, there are some works on the bouncing or cyclic cosmology involving a singularity at the bounce~\cite{Khoury:2001wf, Steinhardt:2001st, Lehners:2008qe}, but we are only interested in the non-singular bounce scenario which can be analyzed solely in the framework of the effective field theory.  Other bouncing or cyclic scenarios can be found in reviews~\cite{Novello:2008ra, Battefeld:2014uga, Brandenberger:2016vhg}.

The rest of the paper is organized as follows. In section~\ref{sec:scenario}, we explain our solution of the bouncing universe based on the positive curvature and the flat potential. In section~\ref{sec:model}, we provide an example model that realizes the sequence of expansion, contraction, and expansion, and also show our numerical results. A possible origin of the positive curvature is discussed in section~\ref{sec:origin}. A variant of the scenario which does not require a negative potential is discussed in section~\ref{sec:cyclic}, and there we find a solution of the cyclic universe.  The last section is devoted to discussion and conclusions.

\section{Expansion, contraction, and expansion again} 
\label{sec:scenario}
In this section, we provide a rough sketch of our bouncing solution and explain
the necessary ingredients. 
Throughout this paper, we assume a homogeneous and isotropic universe
with the FLRW metric,
\begin{align}
\text{d}s^{2}=-\text{d}t^{2}+a^{2}\left(t\right)\left[ \frac { \text{d}{ r }^{ 2 } }{ 
1-\mathcal{K}{ r }^{ 2 } } +{ r }^{ 2 }\left( \text{d}{ \theta  }^{ 2 }+\sin ^{ 2 }{ \theta \text{d} } { \varphi  }^{ 2 } \right)  \right] ,
\end{align}
where $a\left( t \right)$ is the scale factor, 
and $\mathcal{K}$ is the spatial curvature parameter. We assume a closed universe with $\mathcal{K} > 0$ 
for a reason that will become clear shortly.
We consider the Einstein-Hilbert action with a real scalar field $\phi$, 
\begin{align}
S = \int \text{d}^4 x \sqrt{-g} \left( \frac{1}{2} R - \frac{1}{2}g^{\mu\nu} \partial_\mu \phi \partial_\nu \phi - V(\phi) \right),
\end{align}
where $R$ is the Ricci scalar, and $V(\phi)$ is the scalar potential which has a minimum $\phi_{\rm min}$ near the origin. For the moment we assume that $V(\phi_{\rm min})$ is negative, while $V(\phi) > 0$ for the scalar field away from the minimum. This assumption can be relaxed if the spatial curvature is sufficiently large
as we shall see later. 
We adopt the reduced Planck units where $M_{\rm P}= (8\pi G)^{-1/2}\simeq 2.4\times 
10^{18}\ {\rm GeV}$ is set to be unity.

The Friedmann equations  are given by 
\begin{align}
&H^2 = \frac{\rho}{3} - \frac{\mathcal{K}}{a^2}, \\
&\dot{H} = - \frac{1}{2} (\rho+P) + \frac{\mathcal{K}}{a^2}, 
\end{align}
 where the dot denotes the derivative with respect to time, $H=\dot{a}/a$ is the Hubble parameter, 
 and $\rho$ and $P$ are the energy density and pressure, respectively.  In the case of the scalar field, 
 the Friedmann equations and  the equation of motion read
\begin{align}
&H^2  
=\frac{1}{3} \left( \frac{1}{2}\dot{\phi}^2 +V(\phi)  \right) -\frac{\mathcal{K}}{a^2}\, , \label{Friedmann} \\
&\dot{H} 
=-\frac{1}{2}\dot{\phi}^2+ \frac{\mathcal{K}}{a^2}
\, , \label{Hdot} \\
&\ddot { \phi  } +3H\dot { \phi  } +\frac { \partial { V }( \phi )  }{ \partial \phi  } =0\, .\label{EOMphi}
\end{align}

Now let us assume that the universe is initially expanding, $H>0$, and the scalar potential is positive
$V(\phi) >0$. We parametrize the curvature by the density parameter defined by
\begin{align}
\Omega_{\mathcal{K}}& \equiv \frac{\rho_\text{curv}}{\rho_\text{crit}}
=- \frac{3\mathcal{K}/a^2 }{\rho_\text{crit}},
 \end{align} 
 where the critical density is $\rho_\text{crit}  \equiv 3 H^2$.
For the moment we suppose that the curvature is negligible, i.e., 
$|\Omega_\mathcal{K}| \ll 1$.
Then, one can see from eq.~(\ref{Hdot}) that $\dot{H}$ is negative, i.e.~the expansion rate $H$ decreases.  
As the $\phi$ rolls down the potential toward the minimum, it enters the region of the negative potential $V<0$ at a certain point. 
If the kinetic energy and the negative potential balances, the expansion stops $(H=0)$.  
 If the curvature contribution is still subdominant at this point,  eq.~\eqref{Hdot} tells us that $\dot{H}<0$ 
 and $H$ becomes soon negative, i.e.~the universe begins to contract.  
 
Alternatively, the positive curvature can stop the expansion as well and
the universe begins to contract if the kinetic energy is sufficiently large. 
Note that the negative potential is not required in this case. We will return to this possibility
in section~\ref{sec:cyclic}.

During the contraction phase, $\phi$ feels a negative Hubble friction, i.e., anti-friction.
 When $|H|$ is smaller than the mass or characteristic curvature of the scalar potential, 
 the motion of $\phi$ can be approximated by that in the flat space following 
 $\ddot{\phi} + \frac{\partial V}{\partial \phi} \simeq 0$. 
  For example, in the case of a free scalar $V = m^2 \phi^2/2 - \text{const.}$, it behaves as a simple harmonic oscillator
and its oscillation amplitude grows gradually due to the cosmic contraction.  
When $|H|$ becomes comparable to or larger than  the characteristic mass scale,  
the kinetic term generically starts to grow rapidly, $\rho_{\text{kin}} \equiv \frac{1}{2} \dot{\phi}^2 \propto a^{-6}$, 
 following $\ddot{\phi} + 3 H \dot{\phi} \simeq 0$.
In general, it is hard for other components such as radiation, matter,  or curvature to prevent 
the kinetic energy from dominating the universe, as the kinetic energy tends to grow much faster.
Once the universe is dominated by the kinetic energy, 
the universe continues to contract until its size becomes zero: the Big Crunch.
Further details of the cosmological evolution with a negative potential are 
discussed in Refs.~\cite{Linde:2001ae, Felder:2002jk}.

To avoid the Big Crunch, and to make the universe expand again, we need to suppress the growth of the
kinetic energy. To this end, we note that the kinetic energy temporally vanishes at the endpoint in the case of the simple harmonic oscillator. Therefore, if the potential becomes flatter around the
endpoint where $V(\phi)$ is positive, the universe will be dominated by the potential energy for a longer time. Then, the universe will be in the phase of the accelerated contraction  ($\ddot{a}>0$).
In other words, the dynamics governed by the kinetic energy is postponed, and there appears a transient slow-roll regime following $3H\dot{\phi} + \frac{\partial V}{\partial \phi} \simeq 0$.

Although this accelerated contraction does not continue forever since the kinetic energy, however small, 
grows faster than the potential energy,  there is a possibility that the spatial curvature catches up with the potential energy before the kinetic energy becomes relevant. 
This can make the Hubble parameter vanish again, $H=0$. Note that this is possible only if the spatial curvature is positive $\mathcal{K}>0$.  Interestingly, since the kinetic term is subdominant compared to the curvature contribution in eq.~\eqref{Hdot} at this moment,  $\dot{H}>0$ and $H$ becomes soon positive. 
In other words, the universe expands again.  

The above bounce solution resembles the well-known de Sitter bounce solution, 
\begin{align}
a (t)= \sqrt{\frac{3 \mathcal{K}}{\Lambda}} \cosh \left(\sqrt{\frac{\Lambda}{3}} t \right),
\end{align}
 in the presence of the positive spatial curvature $\mathcal{K}$ and 
the positive cosmological constant $\Lambda$. In our case, the positive flat scalar potential plays the role of 
$\Lambda$.  

Generally speaking, the condition of the bounce $\dot{a}=0$ and $\ddot{a}>0$ requires $\rho + 3 P < 0$  at the moment of the bounce. This condition is the same as the accelerated expansion, so inflationary expansion naturally follows the bounce.  In order to continue to the standard inflationary paradigm, the scalar field must find another minimum where the cosmological constant is tiny and positive. 
 We will discuss another possibility to uplift the negative potential later. 

We emphasize that the condition for the bounce solution naturally leads to the subsequent slow-roll inflation. 
If the $e$-folding number is sufficiently 
large, the spatial curvature will be stretched away, and the bounce will be soon forgotten by the attractor nature of inflation.
On the other hand, if the $e$-folding is just enough, we may be able to see the effect of the positive spatial curvature. 
The subsequent cosmological history will be the standard cosmology with slow-roll inflation provided that $\phi$ can find another minimum 
where the potential is small but positive.

\section{Example model}
\label{sec:model}
\begin{figure}[t!] 
        \centering \includegraphics[width=0.9 \columnwidth]{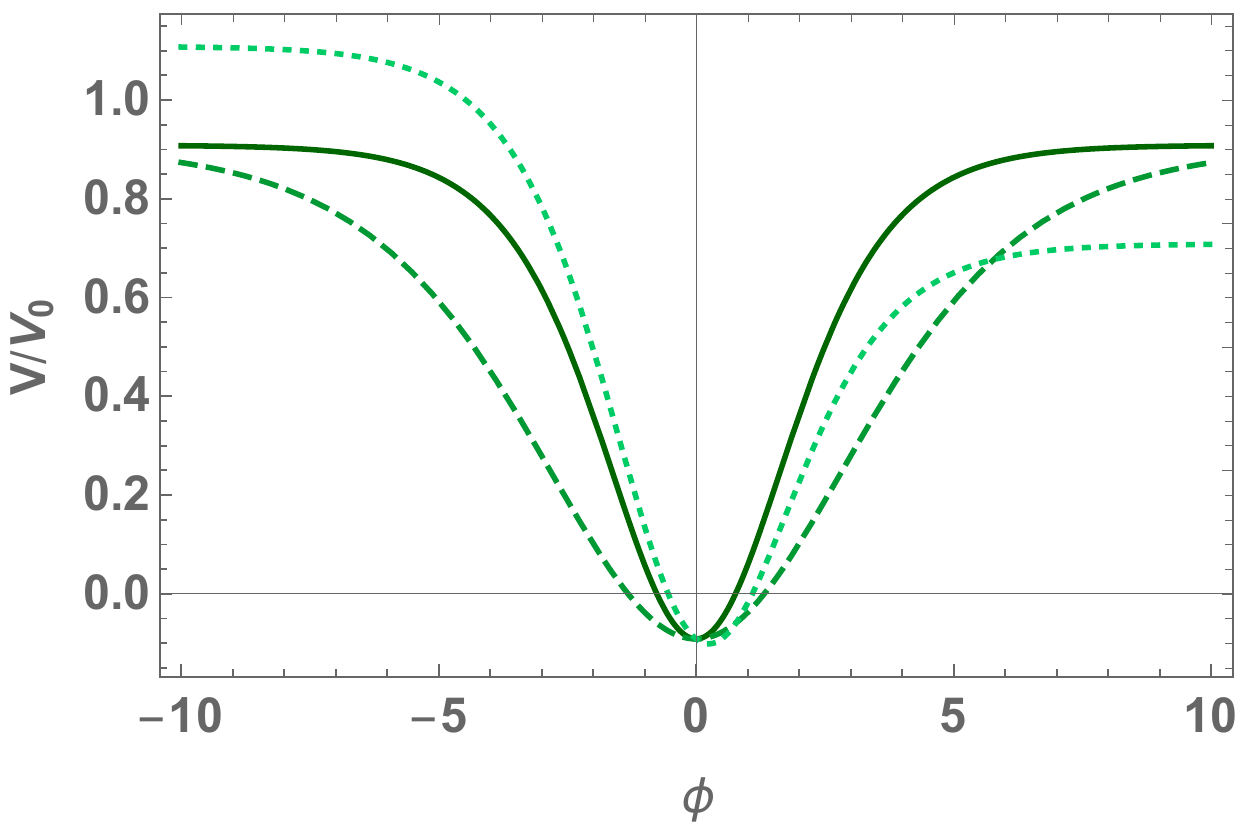}
        \caption{The scalar potential $V$ given by eq.~\eqref{V} with $\gamma = - 0.9143$.  
        Compared to the solid line $(\alpha = 1, \beta = 0)$, the dashed one $(\alpha = 3, \beta = 0)$ has a larger width, 
        and the dotted one $(\alpha = 1, \beta = -0.2)$ is left-right asymmetric. }
        \label{fig:potential}
\end{figure}

To realize the bounce solution described in the previous section,
let us consider the following potential,
\begin{align}
V(\phi) =& V_0 \left( \tanh ^{ 2 }{ \left[ \frac { \phi  }{ \sqrt { 6\alpha  }  }  \right]  } +\beta \tanh { \left[ \frac { \phi  }{ \sqrt { 6\alpha  }  }  \right]  } +\gamma  \right), \label{V}
\end{align}
where $V_0$ is the overall scale of the potential, and 
$\alpha$, $\beta$, and $\gamma$ parametrize  the width, left-right asymmetry and offset of the potential, respectively. 
We consider the following ranges of the parameters,
$\alpha > 0$, $-1 < \beta < 1$, and $-1 < \gamma \leq 0$ in the following. 
The asymptotic flatness of the potential (as well as the appearance of the $\tanh$ function) is
 realized in the context of $\alpha$-attractor~\cite{Ferrara:2013rsa, Kallosh:2013yoa, Galante:2014ifa} 
 or pole inflation~\cite{Galante:2014ifa, Broy:2015qna, Terada:2016nqg}. Similar flattening of the potential can also be realized in the so-called running kinetic inflation~\cite{Takahashi:2010ky,Nakayama:2010kt}.
The potentials for several choices of the
  parameters are shown in Fig.~\ref{fig:potential}.

To realize the bounce solution, the potential must be flat near the endpoint of oscillations where the kinetic energy becomes small.
This can be realized for various choices of the parameters. Suppose that $\phi$ is initially to the left of the potential minimum. 
For a certain value of $\gamma < 0$, 
the universe stops to expand and starts to contract due to the negative potential around the minimum. Then, the kinetic energy starts to increase, and the oscillation
amplitude becomes larger. At a certain point, the endpoint of oscillations approaches the plateau of the potential. It depends
on $\gamma$ how many times the scalar oscillates before it reaches the flat plateau. It is even possible that, for a certain choice of $\gamma$, the scalar reaches the flat potential region on the right without any oscillations after passing through the minimum. 
If the potential on the right plateau is flat enough around the endpoint, the scalar will spend a long time there, and the curvature can catch up with the potential energy, and the bounce occurs. One can similarly adjust $\alpha$ or $\beta$ to realize the bounce solution. 
In the case without any oscillations, larger $\alpha$ and $|\beta|$ ($\beta<0$) make it easier for the scalar field to reach the flat plateau
on the right.

\begin{figure}[tbh] 
     \subcaptionbox{Time evolution of the scale factor.}{
       \includegraphics[width=0.9 \columnwidth]{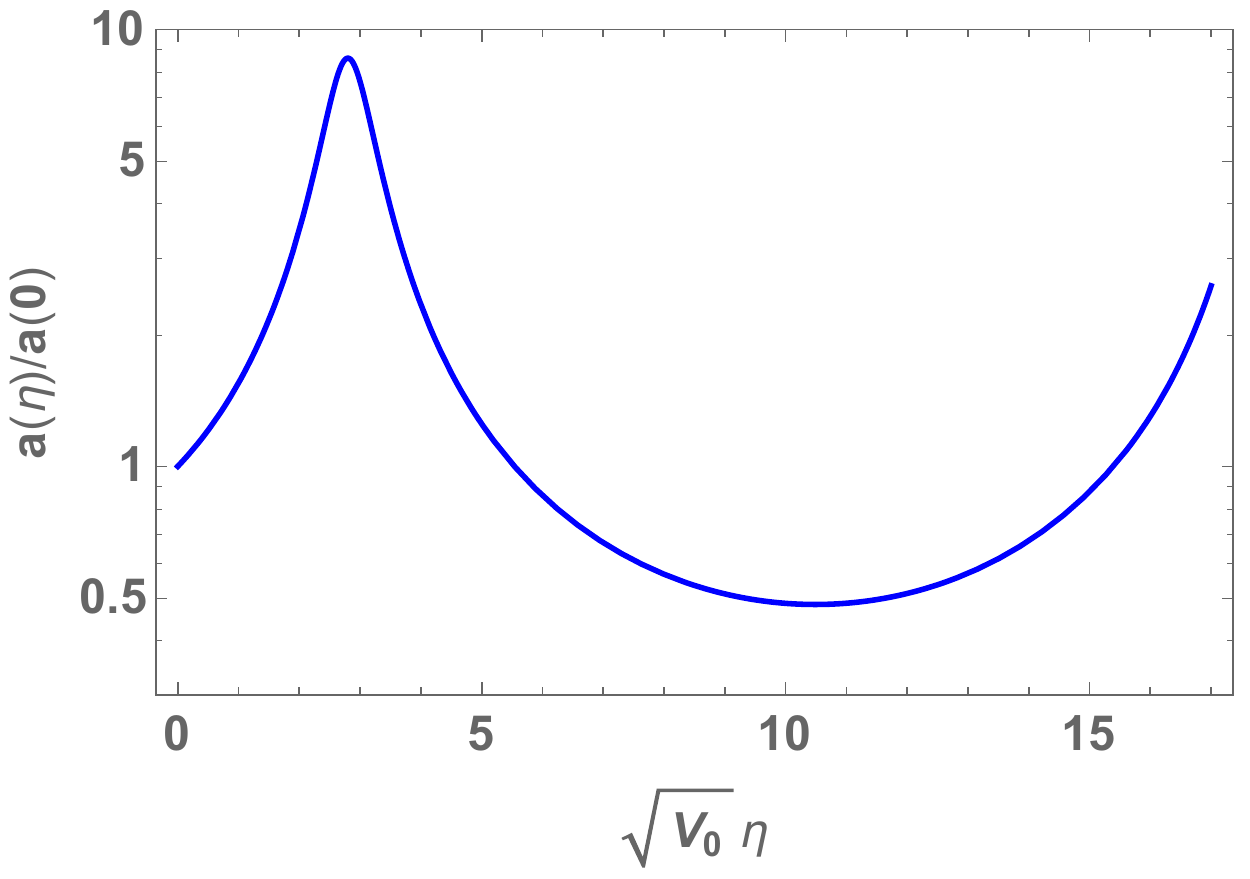}}
       \\
      \subcaptionbox{Time evolution of energy densities.}{
        \includegraphics[width=0.9 \columnwidth]{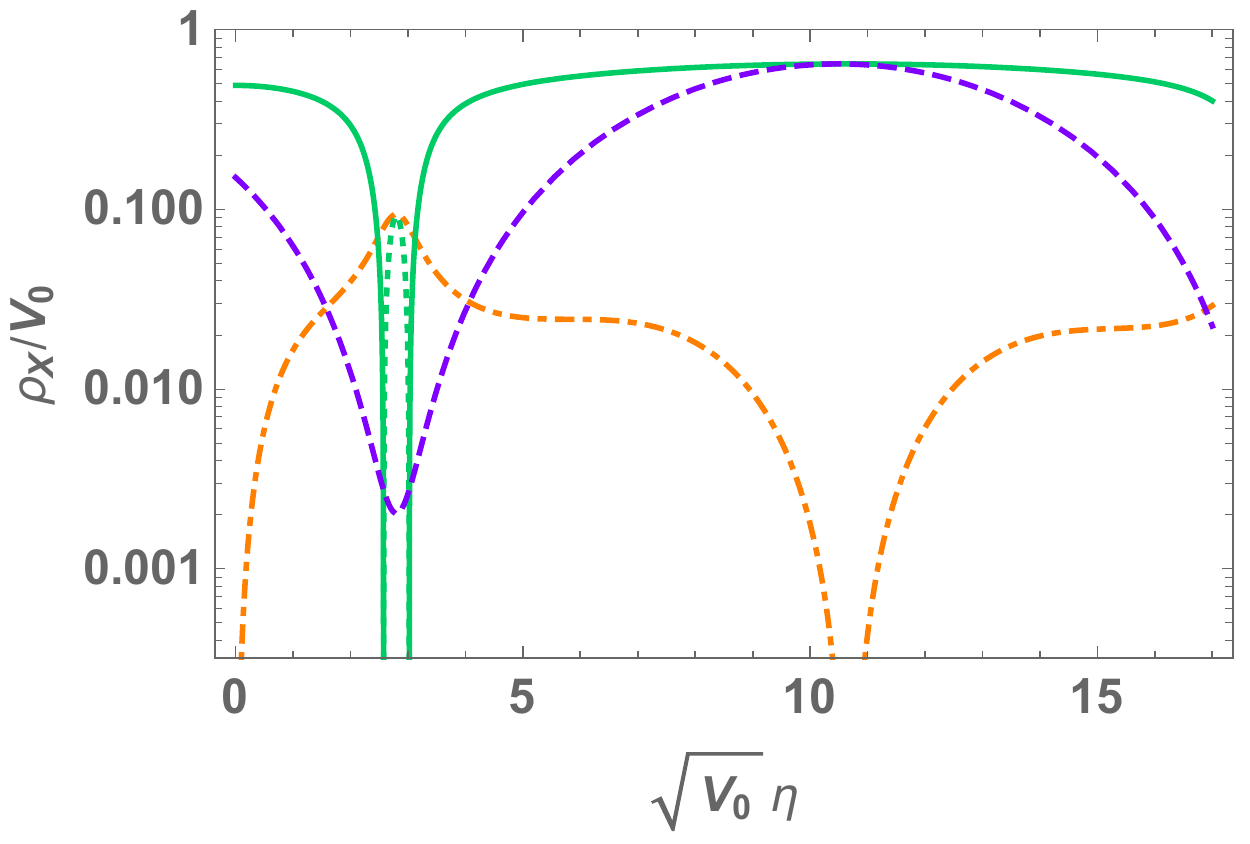}}
        \caption{Bouncing solution with $\alpha = 1$, $\beta=0$, and $\gamma = - 0.09143$.  (top) Time evolution of the scale factor as a function of the conformal time $\eta$. (bottom) Time evolution of the potential energy $V$ (green line; the solid (dotted) one represents $V>0$ ($V<0$)),
the curvature term   $-\rho_\text{curv} = 3 \mathcal{K}/a^2$ (purple dashed line), and the kinetic energy $\dot{\phi}^2/2$ 
(orange dot-dashed line).
}
        \label{fig:numerical_results}
\end{figure}

We have numerically solved the evolution of the universe varying the parameters with
 the initial condition given by
 \begin{align}
\phi(0) =& - \sqrt{6 \alpha},  & & &
\dot{\phi}(0) =& 0, & &\text{and}  & 
\frac{\mathcal{K}}{a(0)^2} = & 0.05 \sqrt{V_0}.
\end{align}
For $\alpha$ of order unity, the cosmic expansion is initially accelerated.
The numerical results for $\alpha = 1$, $\beta=0$, and $\gamma = - 0.09143$ are shown in Fig.~\ref{fig:numerical_results}.  
The top panel shows the evolution of the scale factor as a function of the conformal time, $\eta = \int \text{d}t /a(t)$.
One can see a sequence of the accelerated expansion, accelerated ($\ddot{a}>0$) contraction, and accelerated expansion. 
In this example, $\phi$ does not oscillate before it reaches the right plateau, and it comes back to the potential minimum after
the universe starts to expand again. In order to have sufficiently long inflation after the bounce, we need to modify the potential on the right so that $\phi$ continues to move to the right direction until it finds another potential minimum with almost vanishing cosmological
constant. 

We have also found qualitatively similar bounce solutions for other choices of the parameters, e.g. 
($\alpha = 1$, $\beta=-0.3805885$, and $\gamma = - 0.05$) and ($\alpha = 0.8924$, $\beta = 0$, and $\gamma = - 0.1$).

We have numerically confirmed that the bounce is possible after multiple oscillations around the minimum if we choose
smaller values of $\alpha$ or $|\gamma|$. We note however that, as the number of oscillations increases, the subsequent evolution of the universe becomes more sensitive to the choice of the parameters and exhibits chaotic behavior. 
A similar chaotic evolution was studied in Refs.~\cite{Page:1984qt, Cornish:1997ah, Kamenshchik:1998ue}.

In the above examples without any oscillations, the curvature term is subdominant compared to the kinetic energy or the potential energy when the universe starts to contract. Specifically, $\rho_\text{curv}$ is about two percent of
the kinetic energy and the potential energy around $\eta \approx 3/\sqrt{V_0}$ when $H=0$ in Fig.~\ref{fig:numerical_results}.
We can start with even smaller curvature if we choose parameters such that the kinetic energy becomes more suppressed when the scalar field
reaches the flat part.  Smaller initial curvature implies that the time when the curvature becomes important is delayed, and the length of the contraction period becomes longer.
There is a caveat in such a situation that anisotropic curvature perturbations, in addition to the kinetic term, grow rapidly like $a^{-6}$.  The direction of anisotropy chaotically changes in time, and this phenomenon is known as the BKL instability~\cite{Lifshitz:1963ps, Belinsky:1970ew, Belinski:1973zz} or chaotic mixmaster behavior~\cite{Misner:1969hg} (see e.g.~Ref.~\cite{Erickson:2003zm} for more details).  In non-singular bouncing scenarios, however, such behavior becomes mild~\cite{Ganguly:2017qff}.  Moreover, the initial anisotropy will be suppressed in the situation discussed in section~\ref{sec:origin}.

\section{Possible origin of the positive curvature}
\label{sec:origin}
So far, we have not specified the origin of positive spatial curvature.
A fascinating possibility is the quantum creation of the universe from nothing~\cite{Vilenkin:1982de}. It predicts
the closed universe with $\mathcal{K}>0$.

One introduces
the cosmological wave function $\Psi$ representing the quantum nature of the spacetime 
to estimate the nucleation probability of the universe.
In the so-called mini-superspace approximation, 
we consider only the scale factor $a(t)$
and the homogeneous scalar field $\phi(t)$, and the cosmic wave function $\Psi(a,\phi)$
satisfies the Wheeler-DeWitt equation~\cite{DeWitt:1967yk},
\begin{equation}\label{WDW}
{\cal H}(a,\phi) \Psi(a,\phi) = 0 ,
\end{equation}
with the Hamiltonian given by~\cite{Kiefer:2008sw}
\begin{equation}\label{Hamiltonian}
{\cal H}(a,\phi) = \frac { 1 }{ 12{ a }^{ 2 } } \frac { \partial  }{ \partial a } \left( a\frac { \partial  }{ \partial a }  \right) -\frac { 1 }{ 2{ a }^{ 3 } } \frac { { \partial  }^{ 2 } }{ \partial { \phi  }^{ 2 } }-U(a,\phi) .
\end{equation}
Here we adopt the canonical quantization of the system
to replace the momentum conjugate of $a(t)$ 
with the differential operator $p_a \to -i\frac{\partial}{\partial a}$.
The potential $U(a,\phi)$ is written as
\begin{equation}
U(a,\phi) = { a }^3\left( \frac{3\mathcal{K}}{a^2}-V\left( \phi  \right)  \right) .
\end{equation}
The potential barrier at small values of $a$ suggests that the universe might have emerged 
from nothing ($a=0$) by quantum tunneling.

The universe created in this way at time $t = 0$ has a finite size and obeys the following 
initial conditions~\cite{Vilenkin:1987kf},
\begin{align}
a(0)=&\sqrt { \frac { 3\,\mathcal{K} }{ V\left( \phi  \right)  }  } , &  & &  \ {\dot a}(0)=&0, \nonumber \\ 
\phi(0)=&{\rm const.}, & \text{and}&  &  \ {\dot \phi}(0)=&0.
\label{initial_condition_tunneling}
\end{align}
The nucleation probability will be the highest for the most symmetric configuration, hence the initial inhomogeneity and anisotropy 
are considered to be highly suppressed in this scenario.  
Such homogeneity is advantageous to avoid possible spatial instabilities that are usually present 
when the scalar oscillates in a potential shallower than the quadratic potential.
Even if inhomogeneity and anisotropy grow during such processes, the bounce can occur provided they are small enough~\cite{Anabalon:2012tu, Bramberger:2019zez, Anabalon:2019equ}.

We regard eq.~\eqref{initial_condition_tunneling} as the initial condition of the subsequent classical time evolution.  One can also study how the system dynamically becomes classical by using methods developed in a context of quantum cosmology~\cite{Hartle:2008ng, Battarra:2014xoa,Battarra:2014kga,Lehners:2015sia,Lehners:2015efa}.

There are two different nucleation probabilities of the universe 
due to the ambiguity of the boundary conditions.
Namely, the nucleation probability of the universe for $\phi$ is given by
two forms~\cite{Vilenkin:1987kf},
\begin{equation}
\mathcal{P}_{\rm universe}\left( a, \phi  \right)
\propto \exp\left(\mp \frac{24\pi^2M_{\rm P}^4}{V\left( \phi  \right)}\right)
\end{equation}
where the minus sign corresponds to the tunneling wave function~\cite{Vilenkin:1984wp},
whereas the plus one to the Hartle-Hawking (no-boundary) wave function~\cite{Hartle:1983ai}.
Note that the tunneling wave function predicts that the 
universe nucleates in the highest region of the potential of $\phi$.
Therefore, even without the contraction phase, the nucleated universe
can be smoothly connected to the standard slow-roll inflation.
In contrast, the Hartle-Hawking wave function prefers
the lowest-energy vacuum states of the universe ($V\left( \phi  \right)\rightarrow 0$), and it is often said to
disfavor inflation~\cite{Vilenkin:1987kf,Hawking:1985bk,
Grishchuk:1988br,Barvinsky:1994hx,Page:1997vc}.
We note that, once the universe created from nothing starts to contract and 
experiences the bounce, the energy density can increase by a large amount.
Thus, the potential problem with the Hartle-Hawking wave function can be significantly relaxed.
We will briefly comment on such a possibility in Section~\ref{sec:discussions}.

\section{Cyclic universe}
\label{sec:cyclic}
So far we have mainly focused on the case in which the curvature is subdominant 
when the universe starts to contract, and later it comes to be comparable to the potential energy
to make the universe expand again.
When the curvature is sizable from the very beginning,  on the other hand, it can induce the transitions both from expansion to contraction and from contraction to expansion. 
Interestingly, we find solutions in which expansion and contraction occur alternately,~i.e. the cyclic universe solutions.
In this section, we consider the non-negative potential with $\beta=\gamma = 0$ for simplicity.

\begin{figure}[t!] 
     \subcaptionbox{Evolution of the scale factor.}{
       \includegraphics[width=0.9 \columnwidth]{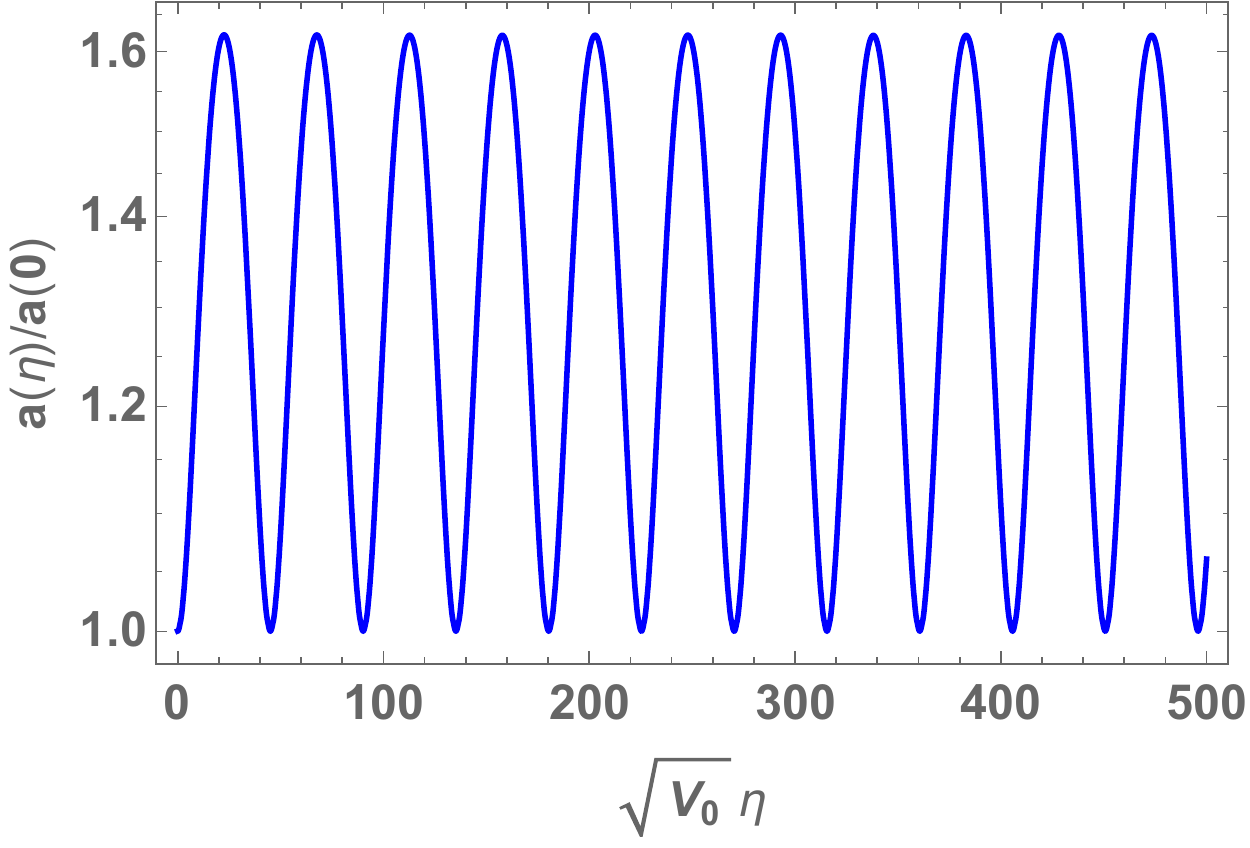}}
       \\
      \subcaptionbox{Magnified view around one of the maxima.\label{sfig:a_cyclic_mag}}{
        \includegraphics[width=0.9 \columnwidth]{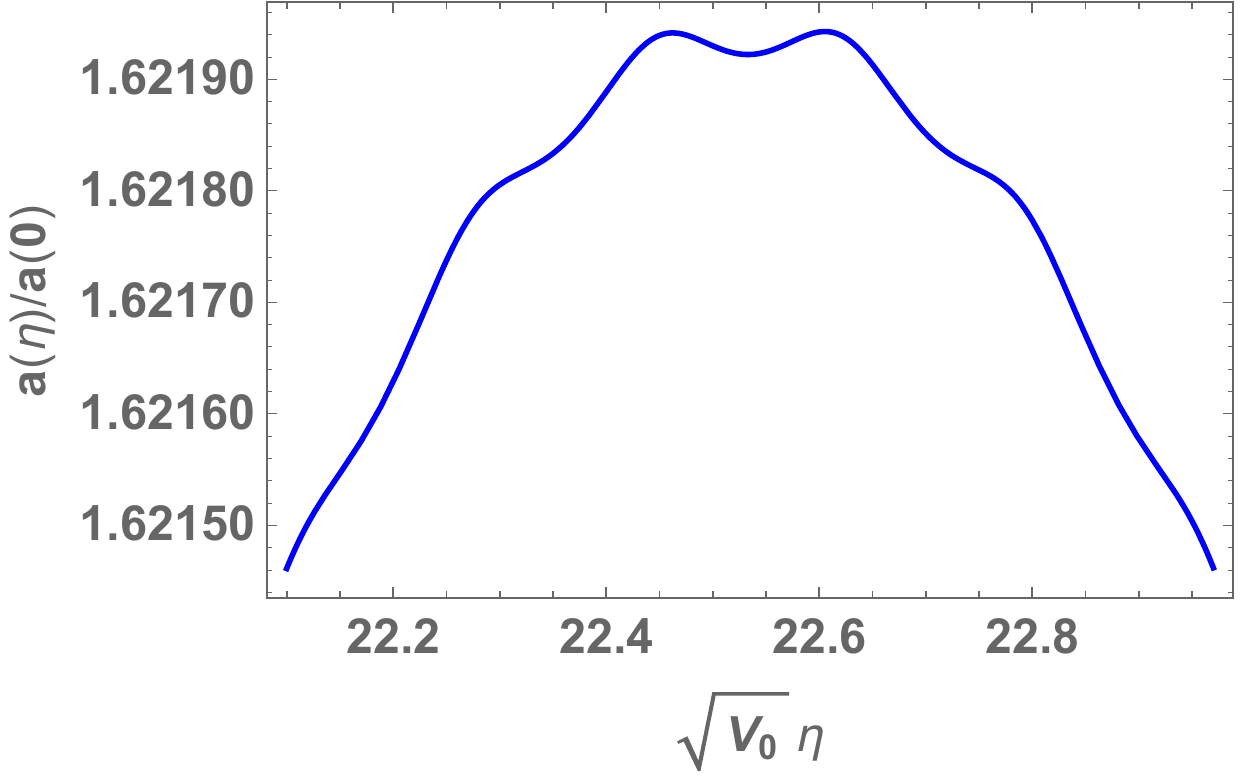}}
        \caption{(top) Time evolution of the scale factor as a function of the conformal time $\eta$ for $\alpha =5 \times  10^{-5}$ and $\beta= \gamma =0$.   (bottom) Magnified view around  $\eta = 22.5/\sqrt{V_0}$ 
        in the top panel. The frequency of the small wiggles corresponds to the mass scale of $\phi$.}
        \label{fig:a_cyclic}
\end{figure}

\begin{figure}[t!] 
     \subcaptionbox{Evolution of energy density components.}{
       \includegraphics[width=0.9 \columnwidth]{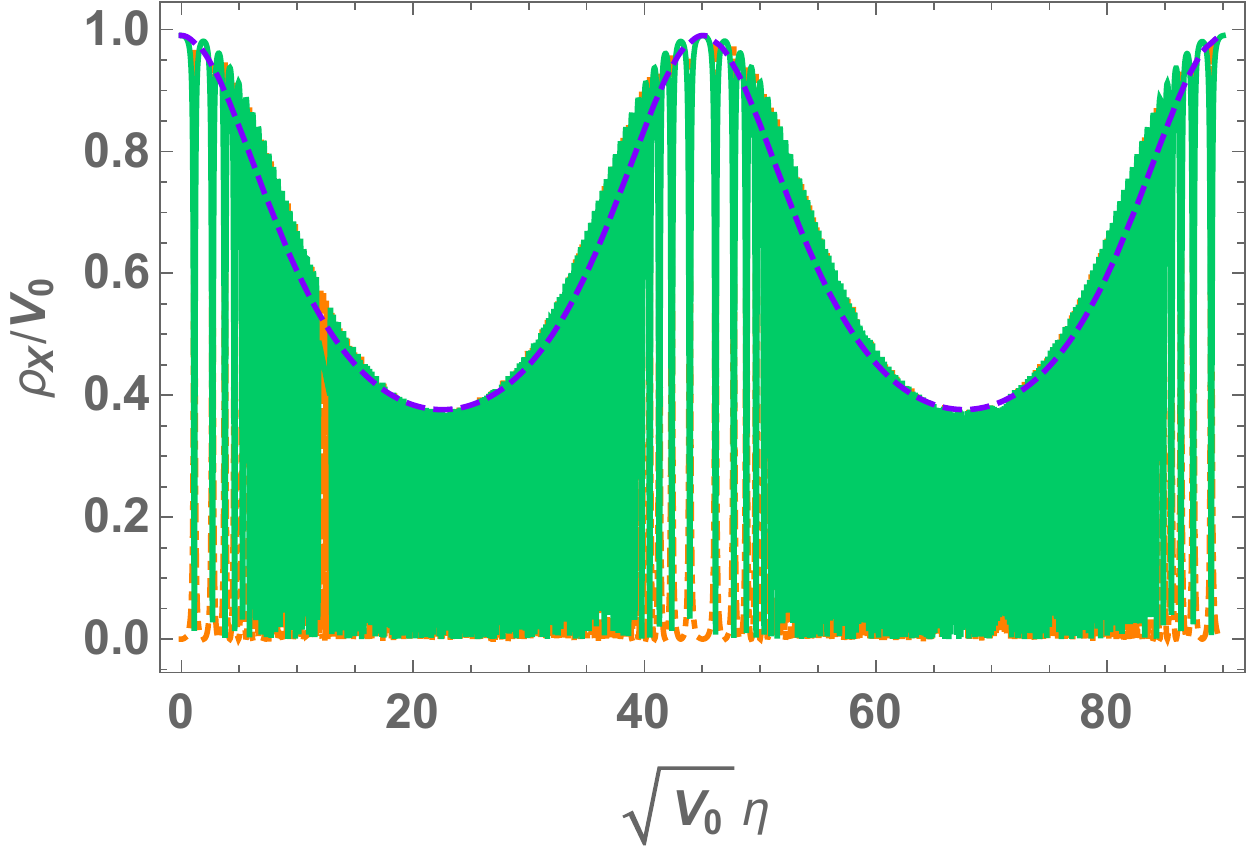}}
       \\
      \subcaptionbox{Magnified view.}{
        \includegraphics[width=0.9 \columnwidth]{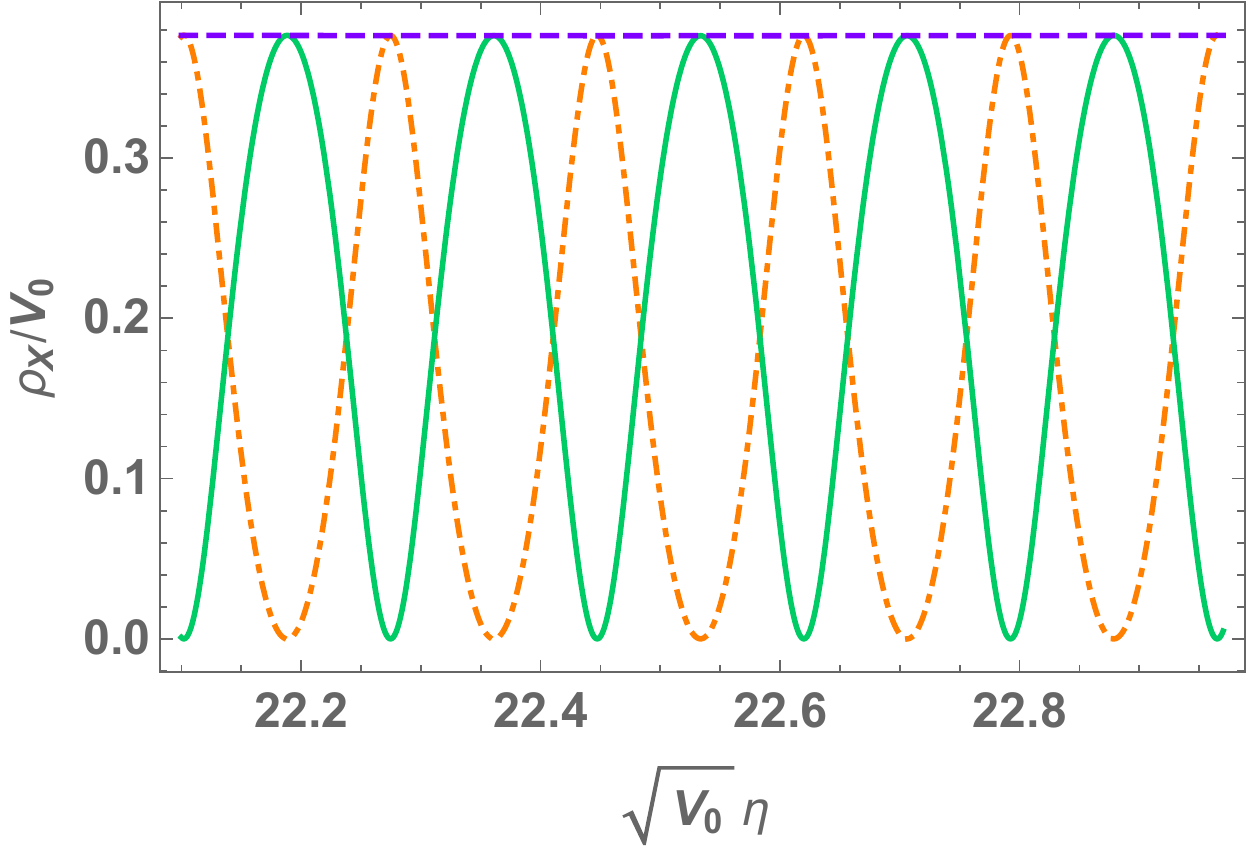}}
        \caption{(top)  
         Time evolution of the potential energy $V$ (green solid),
the curvature term   $-\rho_\text{curv} = 3 \mathcal{K}/a^2$ (purple dashed line), and the kinetic energy $\dot{\phi}^2/2$ 
(orange dot-dashed line) with the same conditions as Fig.~\ref{fig:a_cyclic}. (bottom) Magnified view around $\eta = 22.5/\sqrt{V_0}$ 
        in the top panel.}
        \label{fig:rho_cyclic}
\end{figure}

We find many cyclic universe solutions for sufficiently small values of $\alpha$ for the tunneling initial condition given in eq.~\eqref{initial_condition_tunneling}. The required size of $\alpha$ depends on the initial position of the scalar field. 
One such initial condition is $\mathcal{K}=0.05 V_0$, $\phi(0) = -3 \sqrt{6\alpha}$,  $\dot{\phi}(0)=0$, and $\alpha = 5\times 10^{-5}$.
 We show  in Figs.~\ref{fig:a_cyclic} and \ref{fig:rho_cyclic} the time evolution of the scale factor and the energy densities
for the cyclic universe solution with this choice of the parameters.
 The top panels of the Figures show the cyclic nature of the solution.  Note that the plotted range of $\eta$ is different between the top panels of Figs.~\ref{fig:a_cyclic} and \ref{fig:rho_cyclic}.  In the magnified views in the bottom panels, there are small wiggles corresponding to the oscillations of $\phi$. The frequency of the wiggles is determined by the mass scale of $\phi$. One can also see from
Fig.~\ref{fig:rho_cyclic}  that the curvature contribution is always large in contrast to the case in Fig.~\ref{fig:numerical_results}.  

We may take a coarse-grained view concerning the field oscillations since the cyclic behavior takes place on a much longer timescale as shown in the top panels of the Figures. 
The effective equation of state of the scalar field is obtained by taking an average over oscillations,
$w_{\text{ave}}=\overline{P_\phi/\rho_\phi}$,  where the over-line denotes oscillation average. 
While the universe is expanding, the oscillation amplitude decreases.
For a sufficiently small amplitude, the potential can be approximated by the quadratic one, and
so, $w_{\text{ave}}$ asymptotes to $0$. Then, at a certain point, the curvature becomes important and
the universe stops expanding and starts to contract. While the universe is contracting, the oscillation amplitude increases.
Since our potential becomes flat at large field values, $w_{\text{ave}}$ becomes close to $-1$ if
 the oscillation amplitude is sufficiently large. Then the curvature catches up with the potential energy and the universe bounces.
 The cyclic solution is a result of the interplay between the curvature and the scalar field whose averaged equation of state
 oscillates about $w_{\text{ave}}=-1/3$.

A large number of scalar-field oscillations in a potential shallower than the quadratic one will induce instability of scalar-field fluctuations when the second derivative of the potential becomes non-positive.  
Also, the chaotic mixmaster behavior mentioned at the end of section~\ref{sec:model} may become relevant after many cycles although we expect that such effects can be negligible in the specific example in Fig.~\ref{fig:a_cyclic} because of the not large (order one) ratio of the maximal and minimal values of the scale factor.
Various types of instability in a cyclic setup were studied in Refs.~\cite{Graham:2011nb, Graham:2014pca}.  More generally speaking, the second law of thermodynamics implies that the entropy density increases as the number of cycle increases, so the infinite cycle will be unrealistic~\cite{Tolman:1931zz}. It is not clear whether a large fraction of the universe becomes a viable universe after a large number of cycles.   An attempt is made in Ref.~\cite{Biswas:2008kj} to realize cosmological evolution consistent with observations after many cycles.

\section{Discussion and conclusions } 
\label{sec:discussions}

One of the remaining questions in cosmology is the initial condition of the universe. Did the universe have a beginning? 
If yes, how did the universe begin?
The inflation has moved the initial Big Bang singularity far in the past but does not entirely remove it~\cite{Borde:2001nh}.  In the bouncing universe scenario,
the universe is contracting from the very beginning and starts to expand avoiding any singularities. However, it does not give a satisfactory answer to the question of whether there was a beginning in the universe.

 In this paper, we have first provided a new scenario in which the universe expands, contracts, expands again, and naturally continues to slow-roll inflation based on Einstein gravity with a canonical single scalar field without violating NEC nor encountering any singularity.
We showed that,
for the scenario to be successful, the spatial curvature must be positive and the scalar potential should become 
  flat as the value of the scalar field moves farther away from the potential minimum. 
  Interestingly, our scenario nicely fits into the
creation of the universe from nothing~\cite{Vilenkin:1982de}.
Even if the universe begins to contract due to negative potential or positive curvature after the creation from nothing, 
it can in principle expand again if the curvature term catches up with the scalar energy while
the scalar stays on a flat potential.

It is fair to say that such a nontrivial cosmological history is not realized for an arbitrary choice of the parameters,
and some sort of tuning is necessary. The amount of fine-tuning, however, is not severe at all, 
especially if the initial spatial curvature is relatively large as in the scenario of the creation from nothing. Indeed, 
we have easily found numerous such solutions. However, the fine-tuning becomes severer for a smaller initial spatial 
curvature, and the subsequent evolution of the universe tends to be extremely sensitive to the initial condition and 
the choice of the parameters. The required possible tuning of the parameters may be explained by an anthropic 
argument.
Suppose that the universe is born in a superposition of various states with different shapes of the potential.
Some of the universes may continue to expand and enter the slow-roll inflation regime. Others may start to contract
due to either negative potential or positive spatial curvature, and only those with finely-tuned parameters will
experience the bounce and survive long time for observers to emerge.

We emphasize here that the flat potential required for the bouncing solution keeps the cosmic expansion accelerated after the
bounce. For a certain potential, the slow-roll inflation takes over and lasts for sufficiently large $e$-folds.
For instance, eternal inflation may take place around the maximum of the flat potential. Alternatively, if the $e$-folds are just enough,
we may be able to see the positive spatial curvature in the future observations.
This should be contrasted to a bubble universe scenario in the string landscape which predicts  negative curvature~\cite{Coleman:1980aw, Kachru:2003aw,Guth:2012ww,Kleban:2012ph}.  
Hence, our scenario is in principle falsifiable.  

It is interesting to note that the latest CMB observation may be hinting at the positive spatial curvature:
The Planck bound on the spatial curvature reads $-0.095<\Omega_{\mathcal{K}} < -0.007$ 
 (Planck 2018 TT,TE,EE+lowE 99\% CL), although it is currently consistent with flat universe because, when the lensing and BAO constraints are combined, the limit becomes $\Omega_{\mathcal{K}} = 0.0007 \pm 0.0019$ (68\% confidence level)~\cite{Aghanim:2018eyx}. If future observations confirm the positive curvature, it might be the remnant of the fact that our universe was created from nothing.\footnote{
 We note that positive spatial curvature of order $\Omega_{\cal K} = -0.01$ can relax
the $H_0$ tension~\cite{Riess:2019cxk}.
 } Such a universe might have experienced a contraction phase in a very early age.

We have so far neglected other components such as radiation or matter that may also contribute to the energy density
at the relevant epoch. For our scenario to work, the fraction of such additional components must be sufficiently suppressed. 
This is because, otherwise, they would grow faster than the curvature term.  Once such components dominate
the energy density in the contraction phase, the curvature term cannot compete with them.  Eventually, the kinetic energy tends to
dominate the universe, and the universe will collapse.  Therefore, a finite amount of such extra components limits 
the duration of the contraction phase, and sets a lower bound on the initial spatial curvature for the bounce to occur.

Let us mention an interesting effect induced by an additional light scalar field $\varphi$ 
(such as axions and the Higgs field) which has multiple (metastable) minima with different potential height. We assume that
its initial kinetic energy is sufficiently suppressed. During the contraction phase, the kinetic energy of $\varphi$ 
grows, and if its equation of motion is approximated by $\ddot{\varphi} + 3 H \dot{\varphi} \simeq 0$ with $H<0$,
 its kinetic energy grows like $a^{-6}$. Then, $\varphi$ may go over the potential barrier and reach another minimum with a higher vacuum energy. If $\varphi$ is eventually stabilized at the new minimum after the bounce, the vacuum energy can be dynamically uplifted in the contracting universe~\cite{Graham:2019bfu,Strumia:2019kxg}. 
Then, it may be possible to dynamically set the vacuum energy around the origin of $\phi$ vanishingly small.
In this case, we do not have to modify the potential of $\phi$ (such as eq.~\eqref{V}) to prevent the universe from entering another contracting phase.  After sufficiently long inflation on the plateau,  $\phi$ may return to the 
minimum around the origin.

While we have mainly focused on the case in which the bounce took place in the very early universe,
it is in principle possible that our universe will experience the contraction and bounce in a distant future. This is the
case if the dark energy is due to a quintessence field and the potential becomes negative in the future.
The existence of matter and radiation in the present universe will place a limit on the amount of contraction.
In particular, the contraction must start well after the positive curvature term dominates over matter
for the universe to avoid the Big Crunch.

Before closing, we mention an attractive feature of our scenario.
As mentioned before, the total energy density of the universe increases
during the contraction phase, and it is possible for the universe to
reach an energy scale much higher than the initial state.\footnote{
For instance,
we find a solution in which the energy scale becomes about 100 times higher than the initial value during the contraction phase. 
An example of the parameters leading to such a solution is $\alpha = 10^{-6}$, $\beta=\gamma =0$ and $\mathcal{K}=0.1 V_0$ 
with the initial condition given in eq.~\eqref{initial_condition_tunneling} where $\phi(0)= - 0.1 \sqrt{6 \alpha}$ such that $V(-0.1 \sqrt{6\alpha})
\simeq 0.01 V_0$.}  It may also be possible to modify the potential that allows
an arbitrarily large enhancement. 
There is a class of exponential potentials which allows the ratio $f =V/\rho_\text{kin}$ of the kinetic energy and the potential energy to be constant~\cite{Dabrowski:1995ae, Heard:2002dr}.  The ratio $f$ is related to the equation-of-state parameter $w= P/\rho$ as $w = \frac{1-f}{1+f}$.  Thus, if the solution of $\phi$ around the negative potential
can be smoothly connected to this scaling solution with $w < -1/3$, it allows the bounce at higher energy.  This possibility is attractive especially when the universe is born by the Hartle-Hawking no-boundary proposal~\cite{Hartle:1983ai}, 
which prefers a creation of the universe with lower energy states ($V\left( \phi  \right)\rightarrow 0$) 
and does not predict suitable states for the inflation~\cite{Vilenkin:2002ev}.
However, if the universe experienced the contraction and the energy density 
largely increases after the bounce, the universe might lead to inflation.
We leave a detailed study of such a possibility for future work.

\section*{Acknowledgment}
We thank Masaki Yamada for useful comments on the creation of the universe from nothing. 
This work is supported by JSPS KAKENHI Grant Numbers
JP15H05889 (F.T.), JP15K21733 (F.T.),  JP17H02875 (F.T.), 
JP17H02878 (F.T.) and JP17J00731 (T.T.), by the JSPS Research Fellowship for Young Scientists (T.T.), and by World Premier International Research Center Initiative (WPI Initiative), MEXT, Japan.
\nocite{}
\bibliography{ref}
\end{document}